\documentclass[preprint, superscriptaddress,
showpacs,preprintnumbers,amsmath,amssymb,prb]{revtex4}
\usepackage{graphicx}
\usepackage{amsmath}\usepackage{slashed}

\begin{document}

\thispagestyle{empty}

\title{The van der Waals and Casimir energy of anisotropic atomically thin metallic films}

\author{
G.~L.~Klimchitskaya}
\affiliation{Central Astronomical Observatory at Pulkovo of the Russian Academy of Sciences,
Saint Petersburg,
196140, Russia}
\affiliation{Institute of Physics, Nanotechnology and
Telecommunications, Peter the Great Saint Petersburg
Polytechnic University, St.Petersburg, 195251, Russia}

\author{
V.~M.~Mostepanenko}
\affiliation{Central Astronomical Observatory at Pulkovo of the Russian Academy of Sciences,
Saint Petersburg,
196140, Russia}
\affiliation{Institute of Physics, Nanotechnology and
Telecommunications, Peter the Great Saint Petersburg
Polytechnic University, St.Petersburg, 195251, Russia}

\begin{abstract}
We discuss the van der Waals (Casimir) free energies and pressures of  thin
metallic films, consisting from one to fifteen atomic layers, with regard to
the anisotropy in their dielectric properties. Both
free-standing films and films deposited on a dielectric substrate are considered.
The computations are performed for a Au film and a sapphire substrate.
According to our results, for free-standing Au films consisting of one
and three atomic layers the respective relative error arising from the use
of an isotropic (bulk) dielectric permittivity is equal to 73\% and 37\% for
the van der Waals energy, and 70\% and 35\% for the pressure. We tabulate the
energy and pressure van der Waals coefficients of thin Au films computed
with account of their anisotropy. It is shown that the bulk permittivity of Au
can be used for the films consisting of more than 30 atomic layers, i.e., more
than approximately 7\,nm thickness. The role of relativistic effects is also
investigated and shown to be important even for the films consisting of two or
three layers. The obtained results can find applications in the investigation
of stability of thin films and development of novel nanoscale devices.
\end{abstract}

\maketitle

\section{Introduction}

The van der Waals\cite{1} and Casimir\cite{2} energies and forces are also known
under a generic name of dispersion interactions. They act between closely spaced
material bodies and are caused by the vacuum and thermal fluctuations of the
electromagnetic field whose spectrum is altered by the boundary surfaces.
These are not two different forces. In fact the van der Waals force is
a phenomenon which is  quantum in nature
but  nonrelativistic.
It acts at the shortest separation distances. With increasing
separation, it gradually transforms into the Casimir force, which depends on both the
Planck constant $\hbar$ and the velocity of light $c$. During the last few years the
van der Waals and Casimir interactions attracted much experimental and theoretical
attention in connection with several topical problems of condensed matter physics,
atomic physics, elementary particle physics, and prospective applications in
nanotechnology.\cite{1,2,3,4,5}

Most of the works on the Casimir effect deal with two test bodies separated with either
a vacuum or a liquid-filled gap. There is, however, an important direction in materials
science using heterostructures based on  atomically thin solid films. Such structures
have
already found numerous applications in technology of semiconductor devices, systems for
heterogeneous catalysis and magnetic recording.\cite{6} Thus, in Ref.~[\onlinecite{7}]
it was shown that  ultrathin crystals of MoS${}_2$ consisting from one to six
monolayers exhibit considerably different properties, as compared with the bulk material, and
can be employed as new direct-gap semiconductor. Another example is the use of
atomically thin gold discs for the modification of light at high speeds which has great
potential for nanoscale devices.\cite{8} Taking into account that the Casimir force
between two components of a Si chip has  already been measured,\cite{9} it is important
to investigate the Casimir energies and pressures for  atomically thin material films,
 both the free-standing and deposited on a substrate.

The role of dispersion forces in the stability of thin films has long been discussed in
the literature (see, e.g., the review [\onlinecite{10}]).
The Casimir energy of both the free-standing and deposited on a substrate thin metallic
films was calculated in Refs.~[\onlinecite{11,12}] on the basis of the Lifshitz theory.
In doing so the film metal was considered as an isotropic one. It was described either by the
plasma\cite{11} or by the Drude\cite{12} model. All computations have been performed at
room temperature\cite{11} and at zero temperature.\cite{12}
It is known, however, that for metallic films of thickness less than about $10\lambda_F$,
where $\lambda_F$ is the Fermi wavelength, the boundary effects cannot be neglected,
and the dielectric properties become anisotropic.\cite{13} For example, for Au it holds\cite{14}
$\lambda_F=0.523\,$nm. This means that for Au films containing up to several tens of atomic
layers (such films are of interest for applications mentioned above) the theoretical
description using an isotropic dielectric permittivity is not applicable.

For this reason, in Ref.~[\onlinecite{15}] the Casimir pressure between two thin metallic
films was calculated taking into account an anisotropy of dielectric permittivities by
means of two (in-plane of the film and out-of-plane)l dielectric permittivities.
These permittivities, however, did not allow for the interband transitions of core electrons,
which contribute to the Casimir effect considerably at short separation distances.

Further important progress in the field was achieved\cite{16} in the case of atomically thin films
of Au. On the one hand, in Ref.~[\onlinecite{16}] the more reliable dielectric tensor of Au
films was found within the density functional theory employing the local density
approximation. On the other hand, the interband transitions were taken into account by using the
tabulated optical data\cite{17} for the complex index of refraction for Au extrapolated down to
zero frequency by means of the Drude model. As a result, the Casimir pressure between two
parallel Au films, consisting of several atomic layers, was computed as a function of the
gap width. It was shown\cite{16} that there is an enhancement of the Casimir pressure up to
20\% when the proper anisotropic dielectric permittivities are used, as compared to the
isotropic (bulk) case.

In this paper, we apply the anisotropic dielectric tensor of Ref.~[\onlinecite{16}]
to investigate the Casimir free energy (energy) and pressure of  atomically thin Au films,
both free-standing and
deposited on a sapphire substrate. We calculate the Casimir energy
and pressure in both configurations as functions of the number of atomic layers.
{}From the comparison with similar results obtained using the bulk (isotropic) dielectric
permittivity of Au, we find that the latter can be employed to calculate the Casimir energy of
films containing no less than approximately 30 atomic layers (i.e., for film thicknesses
exceeding approximately 7\,nm). For thinner films the relative deviations of the Casimir
energy calculated using the bulk dielectric properties from the correct results are equal to
73\% and 37\% for the free standing Au films consisting of one and three atomic layers,
respectively. Hence it follows that  film anisotropy leads to significant decrease
in the magnitudes of the Casimir energy and pressure, as compared with computational results
obtained using the bulk dielectric permittivity. This is important for numerous
applications in nanotechnology mentioned above (we recall that in the configuration of
two thin Au films interacting through a vacuum gap an account of film anisotropy enhances
the magnitude of the Casimir force\cite{16}). We have also computed the van der Waals
energy and pressure of  free-standing Au films and Au films deposited on a sapphire
 substrate
in the nonrelativistic limit. A comparison with full computational results using the
Lifshitz theory shows that the relativistic effects play an important role for the thin
films containing at least two or three atomic layers.

The paper is organized as follows. In Sec.~II we present the main expressions of the
Lifshitz theory adapted for the configurations under consideration and calculate
the Casimir energy and pressure for a free-standing Au film with account of its
anisotropy properties. In Sec.~III the Casimir energy and pressure for a film deposited
on a sapphire substrate are computed. Section~IV contains our conclusions and
discussion.

\section{Free-standing gold film}

We consider the free-standing Au film of thickness $a$ in vacuum consisting of $n$
atomic layers, so that $a=nd$, where $d=2.35\,${\AA} is the thickness of one atomic
layer.\cite{16} This film is assumed to be at temperature $T$ in thermal equilibrium with
an environment. The anisotropic properties of film material are described by the
diagonal tensor with the components
$\varepsilon_{xx}^{(0)}(\omega)=\varepsilon_{yy}^{(0)}(\omega)$ and
$\varepsilon_{zz}^{(0)}(\omega)$, i.e., as a uniaxial crystal, where the plane $(x,y)$
is parallel to the film and the $z$ axis is perpendicular to it.

In this case, the Lifshitz formula for the Casimir free energy per unit area can be
found in Refs.~[\onlinecite{18,19}]. Here we follow  modern notations\cite{19}
typical for the scattering theory. Keeping in mind  applications to another configuration
in Sec.~III, we also assume that there are thick isotropic plates (semispaces)
below and above our film which are described by the dielectric permittivities
$\varepsilon^{(-1)}(\omega)$ and $\varepsilon^{(+1)}(\omega)$, respectively.
Then, the Casimir free energy per unit area is given by
\begin{eqnarray}
&&
{\cal F}(a,T)=\frac{k_BT}{2\pi}\sum_{l=0}^{\infty}
{\vphantom{\sum}}^{\prime}\int_{0}^{\infty}k_{\bot}\,dk_{\bot}
\nonumber \\
&&~~~~~~~~~~
\times\left\{\ln\left[1-r_{{\rm TM},\,l}^{(0,+1)}
r_{{\rm TM},\,l}^{(0,-1)}
e^{-2ak_{{\rm TM},\,l}^{(0)}}\right]\right.
\nonumber \\
&&~~~~~~~~~~
+\left.\ln\left[1-r_{{\rm TE},\,l}^{(0,+1)}
r_{{\rm TE},\,l}^{(0,-1)}
e^{-2ak_{{\rm TE},\,l}^{(0)}}\right]\right\}.
\label{eq1}
\end{eqnarray}
\noindent
Here, $k_B$ is the Boltzmann constant,
$k_{\bot}=|\mbox{\boldmath$k$}_{\bot}|$ is the magnitude of the
projection of the wave vector on the plane of the film, and the prime on
the summation sign multiplies the term with $l=0$ by 1/2.
The quantities $k_{{\rm TM,\,TE},\,l}^{(0)}$ contained in the powers of the exponents
in Eq.~(\ref{eq1}) are defined as
\begin{eqnarray}
&&
k_{{\rm TM},\,l}^{(0)}\equiv
k_{\rm TM}^{(0)}(i\xi_l,k_{\bot})=\sqrt{
\frac{\varepsilon_{xx,l}^{(0)}}{\varepsilon_{zz,l}^{(0)}}k_{\bot}^2+
{\varepsilon_{xx,l}^{(0)}}\frac{\xi_l^2}{c^2}},
\nonumber \\
&&
k_{{\rm TE},\,l}^{(0)}\equiv
k_{\rm TE}^{(0)}(i\xi_l,k_{\bot})=\sqrt{
k_{\bot}^2+
{\varepsilon_{xx,l}^{(0)}}\frac{\xi_l^2}{c^2}}
\label{eq2}
\end{eqnarray}
\noindent
for two independent
polarizations of the electromagnetic field, transverse magnetic (TM) and
transverse electric (TE), where
$\xi_l=2\pi k_BTl/\hbar$ with $l=0,\,1,\,2,\,\ldots$ are the
Matsubara frequencies, $\varepsilon_{xx,l}^{(0)}\equiv\varepsilon_{xx}^{(0)}(i\xi_l)$ and
$\varepsilon_{zz,l}^{(0)}\equiv\varepsilon_{zz}^{(0)}(i\xi_l)$.
The reflection coefficients on the interfaces of an anisotropic film and thick isotropic
plates take the form
\begin{eqnarray}
&&
r_{{\rm TM},\,l}^{(0,\pm 1)}\equiv
r_{\rm TM}^{(0,\pm 1)}(i\xi_l,k_{\bot})=\frac{\varepsilon_{l}^{(\pm 1)}
k_{{\rm TM},\,l}^{(0)}-\varepsilon_{xx,l}^{(0)}
k_l^{(\pm 1)}}{\varepsilon_{l}^{(\pm 1)}
k_{{\rm TM},\,l}^{(0)}+\varepsilon_{xx,l}^{(0)}
k_l^{(\pm 1)}},
\nonumber \\
&&
r_{{\rm TE},\,l}^{(0,\pm 1)}\equiv
r_{\rm TE}^{(0,\pm 1)}(i\xi_l,k_{\bot})=\frac{k_{{\rm TE},\,l}^{(0)}-
k_l^{(\pm 1)}}{k_{{\rm TE},\,l}^{(0)}+k_l^{(\pm 1)}},
\label{eq3}
\end{eqnarray}
\noindent
where $\varepsilon_{l}^{(\pm 1)}\equiv\varepsilon^{(\pm 1)}(i\xi_l)$ and
\begin{equation}
k_l^{(\pm 1)}\equiv
k^{(\pm 1)}(i\xi_l,k_{\bot})=\sqrt{
k_{\bot}^2+
{\varepsilon_{l}^{(\pm 1)}}\frac{\xi_l^2}{c^2}}.
\label{eq4}
\end{equation}

In this section we deal with a free-standing Au film in vacuum. Thus,
$\varepsilon_l^{(-1)}=\varepsilon_l^{(+1)}=1$ and from Eqs.~(\ref{eq3})
and (\ref{eq4}) we have
\begin{eqnarray}
&&
r_{{\rm TM},\,l}^{(0,+1)}=r_{{\rm TM},\,l}^{(0,-1)}
=\frac{k_{{\rm TM},\,l}^{(0)}-\varepsilon_{xx,l}^{(0)}
q_l}{k_{{\rm TM},\,l}^{(0)}+\varepsilon_{xx,l}^{(0)}
q_l},
\nonumber \\
&&
r_{{\rm TM},\,l}^{(0,+1)}=r_{{\rm TM},\,l}^{(0,-1)}
=\frac{k_{{\rm TE},\,l}^{(0)}-q_l}{k_{{\rm TE},\,l}^{(0)}+q_l},
\label{eq5}
\end{eqnarray}
\noindent
where
\begin{equation}
q_l\equiv k_l^{(\pm 1)}=\sqrt{k_{\bot}^2+\frac{\xi_l^2}{c^2}}.
\label{eq6}
\end{equation}

The dielectric permittivities of  ultrathin Au films consisting of
$n=1,\,3,\,6,$ and 15 atomic layers were calculated in Ref.~[\onlinecite{16}]
within the density functional theory. They take into account both the effects of
anisotropy and interband transitions of core electrons. In so doing, the tabulated
optical data\cite{17} for the complex index of refraction of Au have been used
extrapolated down to zero frequency by means of the Drude model.

It is well known that there is a problem of great concern in the Lifshitz theory
with respect to this extrapolation. Specifically, theoretical predictions using
the dielectric permittivity extrapolated by the Drude model are found to be excluded
by the experimental data of all precise measurements of the Casimir interaction
between metallic test bodies.\cite{20,21,22,23,24,25,26,26a}
The same measurement data are in a very good agreement with theoretical predictions
using the nondissipative plasma model for an extrapolation of the optical data
to low frequencies.\cite{20,21,22,23,24,25,26,26a}
This is somewhat mysterious if to take into account that the Drude model
allows for the relaxation properties of conduction electrons, which play
a role just at low frequencies, whereas the plasma model disregards the effects
of relaxation.

Fortunately, for the case of an atomically thin Au film considered in this paper,
the above problem does not influence the obtained results. Computations show that
theoretical predictions using the optical data extrapolated to zero frequency by
means of the Drude and the plasma models differ only for film thicknesses
exceeding approximately 30\,nm.\cite{26b}

In Fig.~1, using the results of Fig.~1(a,b) in Ref.~[\onlinecite{16}], we plot the
ratio of the dielectric permittivities
$\varepsilon_{zz}(i\xi)/\varepsilon_{xx}(i\xi)$ as a function of the dimensionless
quantity $\xi/\xi_1$. The lines from bottom to top are for  Au films consisting of
$n=1,\,3,\,6$ and 15 atomic layers, respectively.
By putting $\xi=\xi_l$ one obtains the ratio
$\varepsilon_{zz,l}/\varepsilon_{xx,l}$  as a function of the Matsubara
frequency number $l$.
In an inset, the same information is given  on an
enlarged scale for the first five Matsubara frequencies.
The extrapolation to lower frequencies shows that the ratio
$\varepsilon_{zz}(i\xi)/\varepsilon_{xx}(i\xi)$  goes to zero when the frequency
vanishes along the imaginary frequency axis. As is seen in Fig.~1, for several first
Matsubara frequencies it holds  $\varepsilon_{zz,l}<\varepsilon_{xx,l}$, i.e., there
is an anisotropy of  dielectric properties. The effect of anisotropy decreases
with increasing thickness of the film, as it should be. Thus, for $n=1$ it is
preserved up to $l=160$, whereas for $n=15$ the anisotropy disappears completely
for $l\geq 100$.

Numerical computations of the Casimir free energy have been performed in terms of
the dimensionless Matsubara frequencies $\zeta_l=2a\xi_l/c$. We also introduce two
different dimensionless integration variables in the TM and TE contributions to
Eq.~(\ref{eq1}) by putting $y=2ak_{{\rm TM},\,l}^{(0)}$ and
$y=2ak_{{\rm TE},\,l}^{(0)}$, respectively. Then Eq.~({\ref{eq1}) takes the form
\begin{equation}
{\cal F}(a,T)=-\frac{C_2(a,T)}{a^2},
\label{eq7}
\end{equation}
\noindent
where we have introduced the van der Waals coefficient
\begin{eqnarray}
&&
C_2(a,T)=-\frac{k_BT}{8\pi}\sum_{l=0}^{\infty}
{\vphantom{\sum}}^{\prime}
\int_{\sqrt{\varepsilon_{xx,l}^{(0)}}\,\zeta_l}^{\infty}y\,dy
\nonumber \\
&&~~~~~~~~~~
\times\left\{\frac{\varepsilon_{zz,l}^{(0)}}{\varepsilon_{xx,l}^{(0)}}
\ln\left[1-r_{{\rm TM},\,l}^{(0,+1)}
r_{{\rm TM},\,l}^{(0,-1)}
e^{-y}\right]\right.
\nonumber \\
&&~~~~~~~~~~
+\left.\ln\left[1-r_{{\rm TE},\,l}^{(0,+1)}
r_{{\rm TE},\,l}^{(0,-1)}
e^{-y}\right]
\vphantom{\frac{\varepsilon_{zz,l}^{(0)}}{\varepsilon_{xx,l}^{(0)}}}
\right\}.
\label{eq8}
\end{eqnarray}

In terms of  dimensionless variables the reflection coefficients are given by
\begin{eqnarray}
&&
r_{{\rm TM},\,l}^{(0,\pm 1)}\equiv
r_{\rm TM}^{(0,\pm 1)}(i\zeta_l,y)=
\frac{\varepsilon_{l}^{(\pm 1)}y-\varepsilon_{xx,l}^{(0)}
\sqrt{\frac{\varepsilon_{zz,l}^{(0)}}{\varepsilon_{xx,l}^{(0)}}y^2+
[\varepsilon_{l}^{(\pm 1)}-\varepsilon_{zz,l}^{(0)}]
\zeta_l^2}}{\varepsilon_{l}^{(\pm 1)}y+\varepsilon_{xx,l}^{(0)}
\sqrt{\frac{\varepsilon_{zz,l}^{(0)}}{\varepsilon_{xx,l}^{(0)}}y^2+
[\varepsilon_{l}^{(\pm 1)}-\varepsilon_{zz,l}^{(0)}]
\zeta_l^2}},
\nonumber \\
&&
r_{{\rm TM},\,l}^{(0,\pm 1)}\equiv
r_{\rm TE}^{(0,\pm 1)}(i\zeta_l,y)=\frac{y-\sqrt{y^2+
[\varepsilon_{l}^{(\pm 1)}-\varepsilon_{xx,l}^{(0)}]
\zeta_l^2}}{y+\sqrt{y^2+
[\varepsilon_{l}^{(\pm 1)}-\varepsilon_{xx,l}^{(0)}]
\zeta_l^2}}.
\label{eq9}
\end{eqnarray}
\noindent
Note that the form of $r_{{\rm TM},\,l}^{(0,\pm 1)}$ does not coincide with that in
Ref.~[\onlinecite{19}]. Here, for the sake of convenience, we have introduced another
dimensionless variable $y$ which ensures the common lower integration limit in the
TM and TE contributions to Eq.~(\ref{eq8}). Recall that for a free-standing film
$\varepsilon_l^{(\pm 1)}=1$.

Below we also compute the Casimir (van der Waals) free energy in the nonrelativistic
limit. In this case from Eq.~(\ref{eq9}) we have
\begin{eqnarray}
&&
r_{{\rm nr},\,l}^{(0,\pm 1)}\equiv
r_{{\rm TM},\,l}^{(0,\pm 1)}=
\frac{\varepsilon_{l}^{(\pm 1)}-\sqrt{\varepsilon_{xx,l}^{(0)}
\varepsilon_{zz,l}^{(0)}}}{\varepsilon_{l}^{(\pm 1)}+\sqrt{\varepsilon_{xx,l}^{(0)}
\varepsilon_{zz,l}^{(0)}}},
\nonumber \\
&&
r_{{\rm TM},\,l}^{(0,\pm 1)}=0
\label{eq10}
\end{eqnarray}
\noindent
and Eq.~(\ref{eq8}) reduces to
\begin{equation}
C_2(a,T)=-\frac{k_BT}{8\pi}\sum_{l=0}^{\infty}
{\vphantom{\sum}}^{\prime}
\frac{\varepsilon_{zz,l}^{(0)}}{\varepsilon_{xx,l}^{(0)}}
\int_{0}^{\infty}\!\!\!y\,dy
\ln\left[1-r_{{\rm nr},\,l}^{(0,+1)}
r_{{\rm nr},\,l}^{(0,-1)}
e^{-y}\right].
\label{eq11}
\end{equation}
\noindent
Notice that for the films of small thickness the same computational results are obtained
if we replace the discrete Matsubara frequencies in Eqs.~(\ref{eq8}) and (\ref{eq11})
with the continuous variable $\zeta$ and make a replacement
\begin{equation}
{k_BT}\sum_{l=0}^{\infty}{\vphantom{\sum}}^{\prime}\to
\frac{\hbar c}{4\pi a}\int_{0}^{\infty}\!\!\!d\zeta.
\label{eq12}
\end{equation}
\noindent
This means that for atomically thin films the quantities (\ref{eq8}) and (\ref{eq11})
do not depend on $T$ and have the meaning of the Casimir (van der Waals) energy per
unit area of the film.

In Fig.~2(a) we present the computational results for the van der Waals coefficient
$C_2$ in the free energy (\ref{eq7}) as a function of the number of atomic layers of
a Au film. The values of $C_2$ marked by dots on the solid and dashed lines labeled 1 were
computed using the anisotropic dielectric permittivity of Fig.~1 by fully relativistic
Eq.~(\ref{eq8}) and in the nonrelativistic limit (\ref{eq11}), respectively.
These results refer to the Au films consisting of $n=1,\,3,\,6$ and 15 atomic layers.
For these films the anisotropic dielectric permittivities used in computations have
been found in Ref.~[\onlinecite{16}]. The values of coefficient $C_2$ for the intermediate
values of $n$ were obtained by means of interpolation. The fully relativistic values
of the van der Waals coefficient $C_2$ for the Au films, consisting of $n=1,\,2,\,\ldots ,\,15$
atomic layers are listed in the second column of Table~I.
As is seen in Fig.~2(a) (lines labeled 1), the relativistic effects contribute to the
Casimir energy of the film considerably starting from $n=2$ atomic layers.
Thus, the relative error of the nonrelativistic value of $C_2$, defined as
\begin{equation}
\delta C_{2,{\rm nr}}(n)=\frac{C_{2,{\rm nr}}(n)-C_2(n)}{C_2(n)},
\label{eq13}
\end{equation}
\noindent
quickly increases with increasing $n$: $\delta C_{2,{\rm nr}}=2.0$\%, 6.4\%, 14.5\%,
and 43.7\% for $n=1,\,3,\,6$ and 15, respectively.

The solid and dashed lines labeled 2 in Fig.~2(a) were computed using the bulk
dielectric permittivity of Au by Eq.~(\ref{eq8}) in the relativistic case and
by Eq.~(\ref{eq11}) in the nonrelativistic limit, respectively.
{}From a comparison of the solid lines 1 and 2 in Fig.~2(a) it is seen that the
anisotropy of dielectric properties results in important contribution to the
Casimir energy of the atomically thin Au films. Thus, for films consisting of
$n=1,\,3,\,6$ and 15 atomic layers the ratio of the van der Waals coefficients
obtained using the isotropic and anisotropic dielectric permittivities is equal
to 1.73, 1.37, 1.25, and 1.17, respectively. This means, for instance, that for the
one- and three-layer films the respective deviation of the Casimir energy caused by
a neglect of anisotropy is equal to 73\% and 37\%, respectively.
The influence of film anisotropy practically disappears only for the films consisting of
$n\approx 30$ layers which corresponds to approximately 7\,nm thickness.

Similar results can be obtained for the Casimir (van der Waals) pressure of an
atomically thin metallic film. The Casimir pressure for the configurations under
consideration is found from Eq.~(\ref{eq1})
\begin{eqnarray}
&&
P(a,T)=-\frac{\partial{\cal F}(a,T)}{\partial a}=
-\frac{k_BT}{\pi}\sum_{l=0}^{\infty}
{\vphantom{\sum}}^{\prime}\int_{0}^{\infty}k_{\bot}\,dk_{\bot}
\nonumber \\
&&~~~~~~~~~~
\times\left\{k_{{\rm TM},\,l}^{(0)}\left[
\frac{e^{2ak_{{\rm TM},\,l}^{(0)}}}{r_{{\rm TM},\,l}^{(0,+1)}
r_{{\rm TM},\,l}^{(0,-1)}}-1\right]^{-1}\right.
\nonumber \\
&&~~~~~~~~~~
+\left.k_{{\rm TE},\,l}^{(0)}
\left[
\frac{e^{2ak_{{\rm TE},\,l}^{(0)}}}{r_{{\rm TE},\,l}^{(0,+1)}
r_{{\rm TE},\,l}^{(0,-1)}}-1\right]^{-1}
\right\},
\label{eq14}
\end{eqnarray}
\noindent
where the reflection coefficients are defined in Eq.~(\ref{eq3}).

In terms of the dimensionless variables introduced above, Eq.~(\ref{eq14})
takes the form convenient for numerical computations
\begin{equation}
P(a,T)=-\frac{C_3(a,T)}{a^3},
\label{eq15}
\end{equation}
\noindent
where the van der Waals coefficient $C_3$ is given by
\begin{eqnarray}
&&
C_3(a,T)=\frac{k_BT}{8\pi}\sum_{l=0}^{\infty}
{\vphantom{\sum}}^{\prime}
\int_{\sqrt{\varepsilon_{xx,l}^{(0)}}\,\zeta_l}^{\infty}y^2\,dy
\nonumber \\
&&~~~~~~~~~~
\times\left\{\frac{\varepsilon_{zz,l}^{(0)}}{\varepsilon_{xx,l}^{(0)}}
\left[\frac{e^{y}}{r_{{\rm TM},\,l}^{(0,+1)}
r_{{\rm TM},\,l}^{(0,-1)}}-1\right]^{-1}\right.
\nonumber \\
&&~~~~~~~~~~
+\left.\left[\frac{e^{y}}{r_{{\rm TE},\,l}^{(0,+1)}
r_{{\rm TE},\,l}^{(0,-1)}}-1\right]^{-1}
\vphantom{\frac{\varepsilon_{zz,l}^{(0)}}{\varepsilon_{xx,l}^{(0)}}}
\right\}.
\label{eq16}
\end{eqnarray}
\noindent
Here, the reflection coefficients are presented in Eq.~(\ref{eq9}).

In the nonrelativistic limit Eq.~(\ref{eq16}) is simplified to
\begin{equation}
C_3(a,T)=\frac{k_BT}{8\pi}\sum_{l=0}^{\infty}
{\vphantom{\sum}}^{\prime}
\frac{\varepsilon_{zz,l}^{(0)}}{\varepsilon_{xx,l}^{(0)}}
\int_{0}^{\infty}y^2\,dy
\left[\frac{e^{y}}{r_{{\rm nr},\,l}^{(0,+1)}
r_{{\rm nr},\,l}^{(0,-1)}}-1\right]^{-1}
\label{eq17}
\end{equation}
\noindent
and the reflection coefficients are defined in Eq.~(\ref{eq10}).
Similar to the case of the van der Waals energy, for sufficiently thin films
the substitution (\ref{eq12}) can be made in both Eqs.~(\ref{eq16}) and (\ref{eq17}).
In so doing the computational results do not depend on $T$.

The solid and dashed lines labeled 1 in Fig.~2(b) present the computational results for the
van der Waals coefficient $C_3$ found from Eqs.~(\ref{eq16}) and (\ref{eq17}),
respectively. The computations have been performed for Au films of $n=1,\,3,\,6,$ and 15
atomic layers, using the anisotropic dielectric permittivity  of Ref.~[\onlinecite{16}].
The obtained values of $C_3$ were
 interpolated for the intermediate numbers of layers. The fully relativistic
values of $C_3$ are listed in the third column of Table~I.
The relative error in the nonrelativistic values of $C_3$ defined similar to Eq.~({\ref{eq13})
is equal to $\delta C_{3,{\rm nr}}=0.8$\%, 2.3\%, 5.8\%,
and 19.6\% for Au films consisting of $n=1,\,3,\,6$ and 15 atomic layers, respectively.
One can conclude that for the van der Waals pressure the relativistic effects are somewhat
less than for the van der Waals  energy.

The solid and dashed lines labeled 2 in Fig.~2(b) show the values of the van der Waals
coefficient $C_3$ computed using the bulk dielectric permittivity of Au by
Eqs.~(\ref{eq16}) and (\ref{eq17}), respectively. Just as for the van der Waals energy,
the role of relativistic effect increases with increasing film thickness. The role of
anisotropy in the dielectric properties is also almost the same as for the van der Waals
energy. {}From a comparison of the solid lines 1 and 2 in Fig.~2(b), for
the ratio of the van der Waals coefficients $C_3$,
obtained using the isotropic and anisotropic $\varepsilon$, one finds
1.70, 1.35, 1.22, and 1.13 for $n=1,\,3,\,6$ and 15, respectively.
Thus,  in quantitative determination of the Casimir interaction of the
 atomically thin Au films, it is necessary to take into account an anisotropy of
the dielectric permittivity of Au.

\section{Gold film on a substrate}

Here, we consider an atomically thin Au film deposited on a thick dielectric
substrate. In this case all the above equations apply with $\varepsilon_l^{(+1)}=1$
and $\varepsilon_l^{(-1)}$ equal to the dielectric permittivity of the substrate
material. As an example, we consider the substrate made of Al${}_2$O${}_3$ (sapphire).
The dielectric permittivity of sapphire at the imaginary Matsubara frequencies allows
rather precise analytic representation\cite{27}
\begin{equation}
\varepsilon_l^{(-1)}=1+\frac{C_{\rm IR}\,\omega_{\rm IR}^2}{\omega_{\rm IR}^2+\xi_l^2}
+\frac{C_{\rm UV}\,\omega_{\rm UV}^2}{\omega_{\rm UV}^2+\xi_l^2},
\label{eq18}
\end{equation}
\noindent
where
$C_{\rm UV}=2.072$, $C_{\rm IR}=7.03$, $\omega_{\rm UV}=2.0\times 10^{16}\,$rad/s,
and $\omega_{\rm IR}=1.0\times 10^{14}\,$rad/s.

Numerical computations of the van der Waals coefficient $C_2$ defining the van der Waals
(Casimir) energy (\ref{eq7}) have been performed by Eq.~(\ref{eq8}) in the fully relativistic
case using the anisotropic dielectric permittivity of Ref.~[\onlinecite{16}].
The computational results are shown in Fig.~3(a) as four dots labeled 1 for the Au films
consisting of $n=1,\,3,\,6,$ and 15
atomic layers, respectively. The values of $C_2$ for films consisting of the intermediate numbers
of layers were obtained by means of the interpolation procedure. They are listed in the fourth
column of Table~I. We have also performed respective computations in the nonrelativistic
limit using Eq.~(\ref{eq11}). For an atomically thin Au film deposited on a sapphire
substrate the relative error in the nonrelativistic values of $C_2$ defined in Eq.~({\ref{eq13})
for $n=1,\,3,\,6$ and 15
is equal to $\delta C_{2,{\rm nr}}=3.0$\%, 7.3\%, 15.0\%, and 40.5\%, respectively.
This means that the relativistic effects contribute to the van der Waals  energy
significantly even for very thin films.

For comparison purposes, the line 2 in Fig.~3(a) reproduces the fully relativistic
computational results for the van der Waals coefficient $C_2$ of a free-standing Au film
[in Fig.~2(a) this line is labeled 1]. As is seen in Fig.~3(a), the magnitudes of the
van der Waals energy of atomically thin Au films deposited on a sapphire substrate are
considerably smaller than that of a free-standing film. Thus, the ratio of the respective
van der Waals coefficients $C_2^{\rm Au,\,sa}/C_2^{\rm Au}$ is equal to 0.47, 0.52, 0.54,
and 0.58, for the Au films consisting of $n=1,\,3,\,6$ and 15 atomic layers, respectively.
This means that the deposition on a dielectric substrate considerably decreases
the van der Waals energy
of the atomically thin Au films.

We have also performed numerical computations of the van der Waals coefficient $C_3$ for
 thin Au films deposited on a sapphire substrate. This coefficient determines the van der Waals (Casimir)
pressure (\ref{eq15}). The fully relativistic results computed by Eq.~(\ref{eq16})
using the anisotropic dielectric permittivity of Ref.~[\onlinecite{16}] are shown as
four dots labeled 1 in Fig.~3(b) for  $n=1,\,3,\,6,$ and 15 atomic layers, respectively.
Together with the interpolated values of $C_3$ for other $n$, they are included in the fifth
column of Table~I.
The nonrelativistic values of $C_3$ for a thin Au film deposited on a sapphire
substrate were calculated by Eq.~(\ref{eq17}).
The relative error in the nonrelativistic values of $C_3$
is equal to $\delta C_{3,{\rm nr}}=1.3$\%, 2.9\%, 6.4\%, and 19.0\%
for the Au films consisting of $n=1,\,3,\,6$ and 15 atomic layers, respectively.

The four dots and the line labeled 2 in Fig.~3(b) reproduce the
computational results for the van der Waals coefficient $C_3$ of a free-standing
in vacuum Au film [this line was labeled 1 in Fig.~2(b)].
 As is seen in Fig.~3(b), the deposition on a sapphire substrate
considerably decreases the van der Waals pressure of thin Au films.
Quantitatively, the ratio of the
van der Waals coefficients $C_3^{\rm Au,\,sa}/C_3^{\rm Au}$
in the presence and in the absence of a sapphire substrate is equal to 0.48, 0.52, 0.54,
and 0.57 for the Au films consisting of $n=1,\,3,\,6$ and 15 atomic layers, respectively.
This opens opportunities to control the Casimir  energy and pressure of
the atomically thin metallic films.

\section{Conclusions and discussion}

In the foregoing we have investigated the van der Waals (Casimir) energies and pressures
of metallic films consisting of only a few atomic layers with account of anisotropy of
their dielectric properties. For this purpose, the dielectric tensor of Au obtained in
Ref.~[\onlinecite{16}] using the density functional theory was employed.
Both the cases of  free-standing films and films  deposited on a substrate were considered.
Although metallic films consisting of several atomic layers are not  two-dimensional
systems in a the strict sense, they are somewhat analogous to graphene because are described
by the in-plane and out-of-plane dielectric permittivities (recent progress in calculation
of the Casimir interaction in graphene systems\cite{28,29,30,31,32,33,34,35,36}
resulted in explicit expressions for the dielectric functions of graphene in terms of
the polarization tensor in (2+1)-dimensional space-time\cite{39,40,41,42,43,44,45,46}).
Similar to the case of graphene, we have demonstrated that for atomically thin
metallic films the effect of anisotropy contributes considerably to their van der Waals
(Casimir) energy and pressure and cannot be neglected.

Numerical computations performed for thin Au films demonstrated that their actual
van der Waals energies and pressures are much less than those computed using the bulk
dielectric permittivity. Thus, for the free-standing Au films consisting of one and three
atomic layers the relative error in the van der Waals energy per unit area arising from
the use of bulk dielectric permittivity is equal to 73\% and 37\%, respectively
(similar results hold for the Casimir pressure). This error decreases in magnitude with
increasing number of atomic layers. According to our results, the bulk
(isotropic) dielectric permittivity of Au becomes applicable only for films consisting
of more than 30 atomic layers (i.e., for more than approximately 7\,nm film thickness).
We have also computed the van der Waals (Casimir) energies and pressures for a thin
Au film deposited on a sapphire substrate and tabulated the energy and pressure
van der Waals coefficients in both configurations considered for films consisting
from 1 to 15 atomic layers.

To investigate the role of  relativistic effects in the Casimir energy and pressure
of  atomically thin Au films, we have performed numerical computations in the
nonrelativistic limit. It was shown that for a free-standing Au film the nonrelativistic
results for the van der Waals energy are burdened by the relative error
equal to 2.0\%
and 6.4\% even for one- and three-layer films, respectively. The error increases to
43.7\%  for the Au film consisting of 15 atomic layers (similar errors arise in the
nonrelativistic Casimir energy for a Au film deposited on a sapphire substrate).
The nonrelativistic values of the Casimir pressure for thin films are somewhat more
exact. Thus, for the one- and three-layer free-standing Au films the respective error
is equal to 0.8\% and 2.3\%. For  the film consisting of 15 atomic layers, the error
in the nonrelativistic  Casimir pressure increases up to 19.6\%.

To conclude, we have shown that quantitative description of the van der Waals
(Casimir) energies and pressures of  atomically thin metallic films, both the
free-standing and deposited on substrates, requires an account of anisotropy in
their dielectric properties. Taking into account the wide
application area of such films discussed in
Sec.~I, these results can be useful in  development of novel
heterostructures, semiconductor and nanoscale devices.

\section*{Acknowledgments}

The authors are greatly indebted to Bo~E.~Sernelius for providing
the numerical data of Fig.~1(a,b) in Ref.~[\onlinecite{16}].
We are grateful also to L.~B.~Boinovich for attracting our attention
to the problem of the van der Waals energy of thin films.

%%%%%%%%%%%%%%%%%%%%%%%%%%%%

%%%%%%%%%%%%%%%%%%%%%%%%%%%%
%\end{document}
%%%%%%%%%%%%%%%%
%%%%%%%%%%%%%%%%%
\newpage
%%%%%%%___Table__I__%%%%%%%%%%%%%%%%%%%%
\begingroup
\squeezetable
\begin{table}
\caption{The van der Waals coefficients for the energy per unit area
($C_2$) and pressure ($C_3$) of a Au film consisting of $n$ atomic
layers (column 1) in vacuum (columns 2 and 3) and deposited on a
sapphire substrate (columns 3 and 4).}
\begin{ruledtabular}
\begin{tabular}{rcccc}
 & \multicolumn{2}{c}{ Au film in vacuum} & \multicolumn{2}{c}{ Au film on sapphire} \\
 \cline{2-5}
 $n$ & $C_2$\,(MeV) & $C_3$\,(MeV) & $C_2$\,(MeV) & $C_3$\,(MeV) \\
\hline
1& 42.9 & 86.8 & 20.3 & 41.3\\
2 & 49.0 &100.0 & 24.7 & 50.0\\
3 & 52.2 & 108.5 & 27.0 & 56.3 \\
4 & 53.7 &112.8 & 28.2 & 59.5 \\
5 & 54.1 & 115.1 & 29.0 & 61.8 \\
6 & 53.9 &116.7 & 29.4 & 63.4 \\
7 & 53.6 &117.8 & 29.3 &65.1 \\
8 & 52.8 &118.2 & 29.1 & 66.1 \\
9 & 51.9 & 118.3 & 29.0 & 66.2 \\
10 & 51.0 & 118.0 & 28.9 & 66.5 \\
11 & 50.3 & 117.8 & 28.5 & 66.1 \\
12 & 49.5 & 115.8 & 28.2 & 65.8 \\
13 & 48.6 & 114.2 & 27.8 & 65.2 \\
14 & 47.6 & 112.8 & 27.5 & 64.6 \\
15 & 46.5 & 111.8 & 27.1 & 63.9
%\hline
\end{tabular}
\end{ruledtabular}
\end{table}
\endgroup
%%%%%%%%%%%%%%%%%%%%%%%

%%%%%%%%%%%%%%%%%%%%%%%%%%%%%%%%%%%%%%%%
%%%%%____FIGURE__1___%%%%%%%%%%%%%%%%%%%%%
\begin{figure}[b]
\vspace*{-10cm}
\centerline{\hspace*{1cm}
\includegraphics{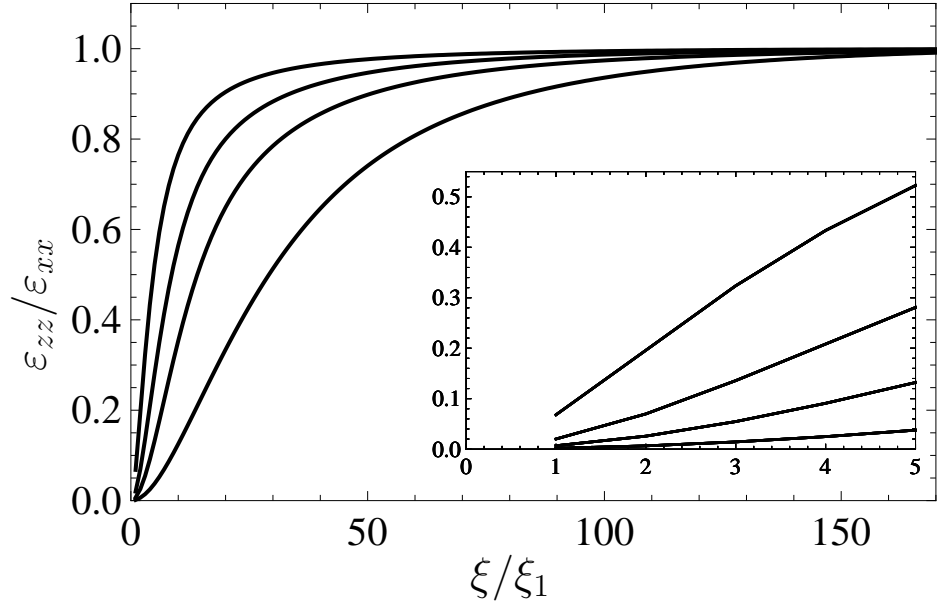}
}
\vspace*{-10cm}
\caption{\label{fg1}
The ratio of the $zz$ to $xx$ components of the dielectric tensor of
the ultrathin Au films is shown as a function of dimensionless imaginary
frequency.
The lines from bottom to top are for the Au films consisting of
$n=1,\,3,\,6$ and 15 atomic layers, respectively.
In an inset the region of small frequencies is shown on an
enlarged scale.
}
\end{figure}
%%%%%%%%%%%%%%
%%%%%____FIGURE__2___%%%%%%%%%%%%%%%%%%%%%
\begin{figure}[b]
\vspace*{-4cm}
\centerline{
\includegraphics{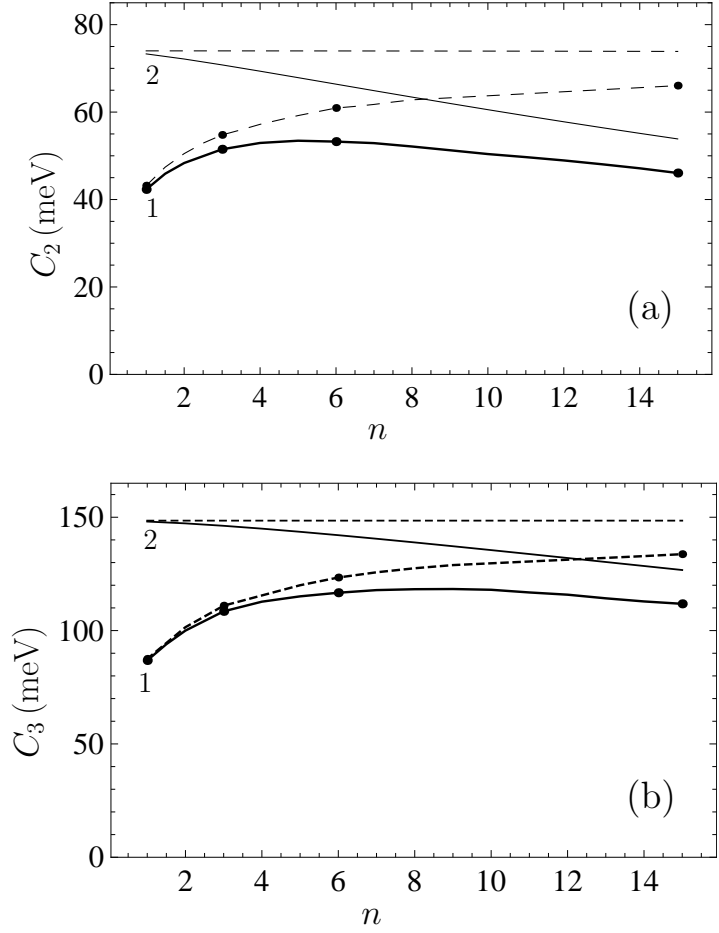}
}
\vspace*{-13cm}
\caption{\label{fg1}
The van der Waals coefficients of the free-standing thin Au film for
(a) the energy per unit area and (b) the pressure are computed using the
anisotropic (the pair of lines labeled 1) and isotropic (the pair of lines
labeled 2) dielectric permittivity of Au are shown as functions of the
number of atomic layers. The solid lines indicate the results of fully
relativistic computations, and the dashed ones are obtained in the
nonrelativistic limit.
}
\end{figure}
%%%%%%%%%%%%%%
%%%%%____FIGURE__3___%%%%%%%%%%%%%%%%%%%%%
\begin{figure}[b]
\vspace*{-4cm}
\centerline{
\includegraphics{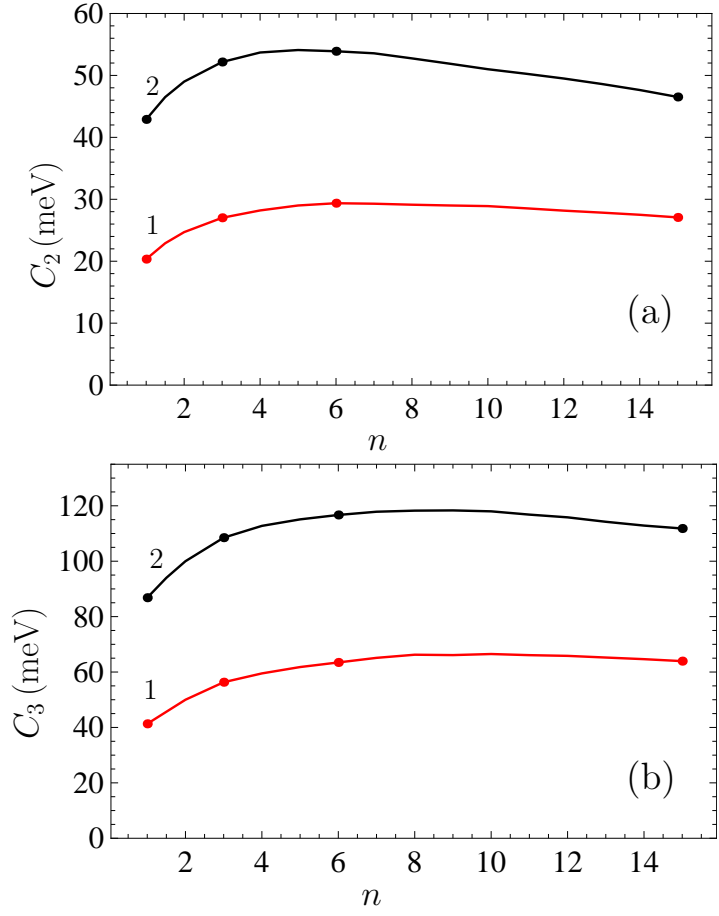}
}
\vspace*{-13cm}
\caption{\label{fg1}(Color online)
The van der Waals coefficients for
(a) the energy per unit area and (b) the pressure computed using the
anisotropic  dielectric permittivity of Au in the configurations of a
thin Au film deposited on a sapphire substrate (the lines labeled 1) and
a free-standing  Au film (the lines labeled 2)
are shown as functions of the
number of atomic layers.
}
\end{figure}
%%%%%%%%%%%%%%
%%%%%%%%%%%%%%%%%%%%%%%%%%%%
\end{document}